\documentclass[aps,prd,twocolumn,superscriptaddress]{revtex4-1}

\usepackage{graphicx}

\usepackage{amsmath}
\usepackage{xfrac}
\usepackage{subfigure}
\usepackage{xcolor}

\newcommand{\be}{\begin{equation}}
\newcommand{\ee}{\end{equation}}
\newcommand{\ba}{\begin{align}}
\newcommand{\ea}{\end{align}}

\newcommand{\x}{\vec{x}}
\newcommand{\p}{\vec{p}}
\newcommand{\pp}{\vec{p}\;\!'}

\newcommand{\q}{\vec{q}}

\newcommand{\lvac}{\langle \, \Omega \, |}
\newcommand{\rvac}{| \, \Omega \, \rangle}
\newcommand{\phia}{\phi^{\alpha}}
\newcommand{\chid}{\chi^{\dagger}}

\newcommand{\phiad}{\phia{}^{\dagger}}

\newcommand{\cG}{G}

\newcommand{\cZ}{\mathcal{Z}}

\renewcommand{\exp}[1]{e^{#1}}
\newcommand{\expi}[2]{e^{-i #1 \cdot #2}}

\begin{document}


\title{Transition of $\rho \rightarrow \pi \gamma$ in Lattice QCD}


\newcommand{\CSSM}{Special Research Centre for the Subatomic Structure
  of Matter (CSSM),\\Department of Physics, University of
  Adelaide, South Australia 5005, Australia} 

\newcommand{\NCI}{National Computational Infrastructure
  (NCI),\\Australian National University, Australian Capital Territory
  0200, Australia} 

\newcommand{\CSIRO}{Digital Productivity Flagship, \\
	CSIRO, 15 College Road, Sandy Bay, TAS 7005, Australia}
	


\author{Benjamin J. Owen}
\affiliation{\CSSM}
\email{benjamin.owen@adelaide.edu.au}

\author{Waseem Kamleh}
\affiliation{\CSSM}

\author{Derek B. Leinweber}
\affiliation{\CSSM}

\author{M. Selim Mahbub}
\affiliation{\CSSM}
\affiliation{\CSIRO}

\author{Benjamin J. Menadue}
\affiliation{\CSSM}
\affiliation{\NCI}


\date{\today}

\begin{abstract}
With the ongoing experimental interest in exploring the excited hadron spectrum, evaluations of the matrix elements describing the formation and decay of such states via radiative processes provide us with an important connection between theory and experiment.  In particular, determinations obtained via the lattice allow for a direct comparison of QCD-expectation with experimental observation.  Here we present the first light quark determination of the $\rho \rightarrow \pi \gamma$ transition form factor from lattice QCD using dynamical quarks.  Using the PACS-CS 2+1 flavour QCD ensembles we are able to obtain results across a range of masses, to the near physical value of $m_{\pi} = 157$~MeV.  An important aspect of our approach is the use of variational methods to isolate the desired QCD eigenstate. For low-lying states, such techniques facilitate the removal of excited state contributions.  In principle the method enables one to consider arbitrary eigenstates. We find our results are in accord with the non-relativistic quark model for heavy masses.  In moving towards the light-quark regime we observe an interesting quark mass dependence, contrary to the quark model expectation.  Comparison of our light-quark result with experimental determinations highlights a significant discrepancy suggesting that disconnected sea-quark loop contributions may play a significant role in fully describing this process.
\end{abstract}


\maketitle

\section{Introduction}
In the search for a complete understanding of the hadron spectrum, hadronic transitions provide a crucial dialogue between experiment and lattice QCD.  Experimental programs such as those at JLAB and Mainz are producing large bodies of high-quality data \cite{Agashe:2014kda} providing valuable insight into the underlying structure and dynamics of hadronic excitations.  Meanwhile, significant effort by the lattice community has seen remarkable progress in understanding and mapping out the hadron spectrum from the perspective of QCD \cite{Thomas:2014dpa,Bulava:2011np}.  Having isolated such states, evaluations of their transition form factors provide crucial insight into their production mechanisms allowing for direct comparison between QCD and the experimental observations.  Another exciting prospect is that lattice determinations may help in the experimental search for as yet unseen QCD eigenstates, such as exotic hadron states being sought by the GlueX experiment \cite{Dudek:2012vr}.

Among the simplest radiative process that can be examined on the lattice is the radiative decay of the rho meson into a pion.  This process is described by a single form factor, $G_{M1}(Q^2)$, and as both states are the lowest lying single particle states in their respective channels, extraction is relatively simple.  Evaluations of this transition were considered early in the course of lattice structure calculations, first by Woloshyn \cite{Woloshyn:1986pk} and later by Lubicz and Crisafulli \cite{Crisafulli:1991pn}.  More recently, our preliminary study \cite{Owen:2012ej} examined both the $\rho$ and $K^*$ transition form factors down to lighter masses ($m_{\pi} \simeq 300$~MeV).  However, all of these studies used quenched gauge field configurations.  Currently the only evaluations in full QCD are those of Edwards \cite{Edwards:2004xj} and Shultz \cite{Shultz:2015pfa} in which this transition along with a range of other elastic and transitions  form factors were considered.  However, the parameters considered were far from physical, particularly the large pion mass.  There is thus a need to examine this transition at light quark masses using dynamical quarks. 

Recently the issue of excited states has received renewed focus \cite{Dinter:2011sg,Lin:2012ev,Alexandrou:2010cm} as lattice determinations of several quantities, most notably the nucleon axial charge, $g_A$, and quark momentum fraction, $\langle x \rangle$, continue to differ systematically from the experimental results despite operating at near physical quark masses and large lattice volumes. In Refs. \cite{Owen:2012ts,Owen:2015gva}, we presented a framework for utilising variational techniques for the extraction of hadronic matrix elements. For both the calculation of $g_A$ \cite{Owen:2012ts} and the electromagnetic form factors of the $\pi$ and $\rho$ mesons \cite{Owen:2015gva}, we found that the use of correlation matrix methods led to significantly improved ground state dominance over the standard single smeared source approach. This in turn allowed for the use of earlier current insertion times and subsequently consider earlier fit windows resulting in improved statistical uncertainties. Furthermore, within Ref. \cite{Owen:2015gva} we were able to examine the corresponding matrix elements for both excited pi and rho meson states.  The same techniques were utilised by Shultz et al. \cite{Shultz:2015pfa} to examine the transition form factors for the lightest $\pi$ and $\rho$ excited states.  In this work we will make use of such techniques in our determination of the $\rho \rightarrow \pi \gamma$ transition form factors across a wide range of quark masses including the light-quark regime.

The remainder of this investigation is organized as follows. In section 2 we present a brief summary of the use of variational methods in the evaluation of transition matrix elements. Section 3 outlines how we intend to extract the transition form factor, $G_{M1}(Q^2)$ and finally a summary of calculation details in Section 4. Our results are discussed in Section 5 followed by concluding remarks in Section 6.

\section{Correlation matrix methods for hadron transitions}

The use of variational techniques for the calculation of hadronic matrix elements has been presented in Refs. \cite{Owen:2012ts,Owen:2015gva} and here we present a brief outline to define our procedure and notation. Similar prescriptions have been explored in Refs. \cite{Dudek:2006ej,Dudek:2009kk,Bulava:2011yz,Lewis:2011ti,Maurer:2012sf,Blossier:2013qma}. The variational method \cite{Michael:1985ne,Luscher:1990ck} is an approach to generating a set of ideal operators, $\lbrace \phi^{\alpha} \rbrace$, constructed to have overlap with an individual eigenstate
\be
	\label{eq:diagCond}
	\lvac \, \phia  \, | \, \beta, p \, \rangle \propto \delta^{\alpha \beta} \, .
\ee
This is achieved by choosing an existing basis of operators, $\lbrace \chi_i \rbrace$, and constructing these idealised operators as linear combinations
\be
	\label{eq:superpos}
	\phia(x) = \sum_{i} v^{\alpha}_{i} \, \chi_{i}(x), \hspace{15pt} \phiad(x) = \sum_{j} \chid_{j}(x) \, u^{\alpha}_{j} \, .
\ee
Starting from the matrix of two-point correlation functions
\[
	\cG_{ij}(\p, t) = \sum_{\x} \expi{\p}{\x} \, \lvac \, \chi_i(x) \; \chid_j(0) \, \rvac \, ,
\]
one can show that both $v^{\alpha}_{i} \, \cG_{ij}(\p, t + \delta t)$ and $\cG_{ij}(\p, t+ \delta t) \, u^{\alpha}_{j}$ provide a recurrence relation from which the necessary vectors, $v^{\alpha}_{i}$ and $u^{\alpha}_{j}$, are the solutions of the following generalised eigenvalue equations
\begin{subequations}
\begin{align}
	\label{eq:LeftEV}
	v^{\alpha}_{i} \, \cG_{ij}(\p, t_0 + \delta t) = & \, \exp{-E_{\alpha}(\p) \, \delta t} \, v^{\alpha}_{i} \, \cG_{ij}(\p, t_0) \\
	\label{eq:RightEV}
	\cG_{ij}(\p, t_0 + \delta t) \, u^{\alpha}_{j} = & \, \exp{-E_{\alpha}(\p) \, \delta t} \, \cG_{ij}(\p, t_0) \, u^{\alpha}_{j} \, .
\end{align}
\end{subequations}
With the optimised operator for the state $| \, \alpha, p \, \rangle$ suitably defined, one can construct the corresponding correlator by projecting the matrix of correlators
\[
	\cG(\p, t; \alpha)= (v(\p))^{\alpha}_{i} \: \cG_{ij}(\p, t) \: (u(\p))^{\alpha}_{j} \, .
\]
Here we note that the correlators that feed into Eq.~\eqref{eq:LeftEV} and \eqref{eq:RightEV} are functions of the 3-momentum, $\p$. Therefore it is necessary to evaluate eigenvectors for each momentum considered. This is of particular relevance to three-point correlators where the incoming and outgoing momenta in general differ.

To obtain the three-point correlator necessary for describing a particular transition, it is a simple matter of projecting the relevant states with the corresponding eigenvectors for source and sink.  In order to do this, one first evaluates the matrix of three-point correlation functions required for the transition in question
\begin{align}
	\label{threept}
	(\cG^{a \rightarrow b}_{{\cal O}})_{ij}(\pp, \p, t_2&, t_1) = \sum_{\x_2, \x_1} \expi{\pp}{(\x_2-\x_1)} \expi{\p}{\x_1} \nonumber \\
	& \times \lvac \, \chi_{b,i}(x_2) \: {{\cal O}}(x_1) \: \chid_{a,j}(0) \, \rvac \, ,
\end{align}
where ${\cal O}$ is the current operator through which the transition takes place and the labels $a$ and $b$ highlight that the source and sink operators can correspond to different $J^{PC}$. In this situation, one must construct ideal operators for each set of quantum numbers. By performing the variational analysis with the bases $\lbrace \, \chi_{a,i} \, \rbrace$ and $\lbrace \, \chi_{b,i} \, \rbrace$, we obtain two sets of eigenvectors: $(v_a)^{\alpha}_{i}$, $(u_a)^{\alpha}_{j}$ and $(v_b)^{\beta}_{i}$, $(u_b)^{\beta}_{j}$ which form optimised operators of type $a$ and $b$ respectively. It is then a simple matter of projecting the necessary eigenvectors onto the matrix of three-point functions, where care is taken to ensure that the projection is done with the correct momenta and operator for source and sink
\begin{align*}
	\cG^{a \rightarrow b}_{{\cal O}}(\pp, \p, t_2&, t_1; \beta,\alpha) \equiv \\
	& (v_b(\pp))^{\beta}_{i} \, (\cG^{a \rightarrow b}_{{\cal O}})_{ij}(\pp, \p, t_2, t_1) \, (u_a(\p))^{\alpha}_{j} \, .
\end{align*}

Having acquired the projected two and three-point correlation functions, determination of matrix elements then follows in the standard way through the construction of a suitable ratio. For transitions, we choose to work with a modified form of the ratio presented in Ref. \cite{Hedditch:2007ex}
\begin{align}
	\label{eq:Ratio}
	R(p'&, p; \beta, \alpha) = \nonumber \\
 &\sqrt{ \frac{ \langle \cG^{a \rightarrow b}_{{\cal O}}(\pp, \p, t_2, t_1; \beta, \alpha) \rangle \, \langle \cG^{b \rightarrow a}_{{\cal O}}(\p, \pp, t_2, t_1; \alpha, \beta) \rangle }{ \langle \cG_{a}(\p, t_2; \alpha) \rangle \, \langle \cG_{b}(\pp, t_2; \beta) \rangle } } \, .
\end{align}
Though this requires the evaluation of both $+\q$ and $-\q$ sequential source technique (SST) propagators \cite{Bernard:1986,Bernard:1984}, if one wishes to use both $\lbrace U \rbrace$ and $\lbrace U^* \rbrace$ configurations as was done in Refs. \citep{Owen:2015gva,Boinepalli:2006xd} and will be done here, then these are required regardless of the choice of ratio \cite{Draper:1988xv}.

\section{$\rho \rightarrow \pi \gamma$ transition Form Factors}

Having outlined the method for obtaining the projected correlators and the corresponding ratios, we now consider how to isolate the form factor describing the radiative decay of the rho meson.  In order to do so we make the identification that $a = \pi$, $b = \rho$ and ${\cal O} = J^{\mu}$ the electromagnetic current.  For this discussion we shall work with the standard Minkowski metric.  Starting from the projected three-point correlation function
\begin{align}
	\label{eq:threept}
	(\cG_{\rho \rightarrow \pi\gamma})^{\mu}_{\nu}(\p, \pp&, t_2, t_1; \,\alpha, \beta) = \sum_{\x_2, \x_1} \expi{\p}{(\x_2-\x_1)} \expi{\pp}{\x_1} \nonumber \\
	&\times \lvac \, \phi^{\alpha, \p}_{\pi}(x_2) \, J^{\mu}(x_1) \, \phi^{\beta, \pp}_{\rho, \nu}{}^\dagger(0) \, \rvac \, ,
\end{align}
we begin by inserting completeness identities between our operators in order to obtain the desired matrix element, as well as operator overlap factors. This matrix element, which describes the transition, can be parametrised by a single form factor $G_{M1}$
\begin{align}
\label{eq:vertex}
\langle \, \pi_{\alpha}(\p) \, | \, &J^{\mu}(0) \, | \, \rho_{\beta}(\pp, s') \, \rangle = \frac{1}{2\,\sqrt{E_{\pi_{\alpha}}(\p)\, E_{\rho_{\beta}}(\pp)}} \nonumber \\
	 & \times \left( \frac{-ie}{m_{\rho_{\beta}}} \right) G_{M1}(Q^2) \, \varepsilon^{\mu\delta\sigma\tau} \, p'_{\delta} \, p_{\sigma} \, \epsilon^{\beta}_{\tau}(p',s') \, ,
\end{align}
where we note that our choice of normalisation is that of Ref. \cite{Woloshyn:1986pk}. Furthermore we note the pion overlap can be expressed as
\[
	\lvac \, \phi_{\pi}^{\alpha,\p}(0) \, | \, \pi_{\beta}(\p) \, \rangle = \frac{\delta^{\alpha\beta}}{\sqrt{2\,E_{\pi_\alpha}(\p)}} \, \cZ^{\alpha}_{\pi}(\p)
\]
and similarly the rho meson overlap
\[
	\lvac \, \phi_{\rho, \nu}^{\alpha, \p}(0) \, | \, \rho_{\beta}(\p, s) \, \rangle = \frac{\delta^{\alpha\beta}}{\sqrt{2\,E_{\rho_\alpha}(\p)}} \,  \epsilon^{\alpha}_{\nu}(p,s) \, \cZ^{\alpha}_{\rho}(\p)
\]
where $\epsilon(p,s)$ is a spin polarisation vector which satisfies
\[
	\sum_s \epsilon_{\mu}(p,s) \, \epsilon^*_{\nu}(p,s) = - \left( g_{\mu\nu} - \frac{p_{\mu}p_{\nu}}{m_{\rho_{\alpha}}} \right) \, .
\]
Substituting these expressions into Eq.~\eqref{eq:threept} and making use of the spin-sum identity we arrive at
\begin{align*}
	(\cG_{\rho \rightarrow \pi\gamma}&)^{\mu}_{\nu}(\p, \pp, t_2, t_1; \alpha, \beta) = \\
	& \frac{\exp{-E_{\pi_{\alpha}}(\p)\,(t_2-t_1)} \exp{-E_{\rho_{\beta}}(\pp)\,t_1}}{4\,E_{\pi_\alpha}(\p)\,E_{\rho_\beta}(\pp)} \, \cZ^{\alpha}_{\pi}(\p) \, \cZ^{\beta}_{\rho}{}^\dagger(\pp) \\
	& \times \left( \frac{+ie}{m_{\rho_{\beta}}} \right) \, G_{M1}(Q^2) \, \varepsilon^{\mu\delta\sigma\tau} \, p'_{\delta} \, p_{\sigma} \, \left( g_{\tau\nu} - \frac{p'_{\tau}p'_{\nu}}{m_{\rho_\beta}^2} \right) \, .
\end{align*}
Taking the Hermitian conjugate of Eq.~\eqref{eq:vertex}, we can obtain the time-reversed matrix element necessary for $\cG_{\pi\gamma \rightarrow \rho}$. Following the same procedure we arrive at the corresponding expression
\begin{align*}
	(\cG_{\pi\gamma \rightarrow \rho}&)^{\mu}_{\nu}(\pp, \p, t_2, t_1; \beta, \alpha) = \\
	& \frac{\exp{-E_{\rho_{\beta}}(\pp)\,(t_2-t_1)} \exp{-E_{\pi_{\alpha}}(\p)\,t_1}}{4\,E_{\pi_\alpha}(\p)\,E_{\rho_\beta}(\pp)} \, \cZ^{\beta}_{\rho}(\pp) \, \cZ^{\alpha}_{\pi}{}^\dagger(\p) \\
	& \times \left( \frac{-ie}{m_{\rho_{\beta}}} \right) \, G_{M1}(Q^2) \, \varepsilon^{\mu\delta\sigma\tau} \, p'_{\delta} \, p_{\sigma} \, \left( g_{\tau\nu} - \frac{p'_{\tau}p'_{\nu}}{m_{\rho_\beta}^2} \right) \, .
\end{align*}
In order to cancel out the time dependence and the overlap factors, $\cZ$, we require the projected two-point functions
\begin{align*}
	\cG_{\pi}(\p, t_2; \alpha) &= \sum_{\x_2} \expi{\p}{\x_2} \lvac \, \phi^{\alpha, \p}_{\pi}(x_2) \, \phi^{\alpha, \p}_{\pi}{}^\dagger \,(0) \, \rvac \\
	(\cG_{\rho})_{\nu\nu}(\pp, t_2; \beta) &= \sum_{\x_2} \expi{\pp}{\x_2} \lvac \, \phi^{\beta, \pp}_{\rho, \nu}(x_2) \, \phi^{\beta, \pp}_{\rho,\nu}{}^\dagger \,(0) \, \rvac \, ,
\end{align*}
where we note that repeated indices are not summed over. Again applying completeness and replacing spin-sums, these expressions reduce to
\begin{align*}
	\cG_{\pi}(\p, t_2) &= \frac{\exp{-E_{\pi_{\alpha}}(\p)t_2}}{2E_{\pi_\alpha}(\p)} \, \cZ^{\alpha}_{\pi}(\p) \, \cZ^{\alpha}_{\pi}{}^\dagger(\p) \\
	(\cG_{\rho})_{\nu\nu}(\pp, t_2) &= - \frac{\exp{-E_{\rho_{\beta}}(\pp)t_2}}{2E_{\rho_\beta}(\pp)} \, \cZ^{\beta}_{\rho}(\pp) \cZ^{\beta}_{\rho}{}^\dagger(\pp) \\
	& \qquad\qquad \times \left( g_{\nu\nu} - \frac{p_{\nu}p_{\nu}}{m_{\rho_\beta}^2} \right) \, .
\end{align*}
Finally, substituting in the explicit form of both two and three-point correlation functions into Eq.~\eqref{eq:Ratio}, we obtain the expression which directly relates our ratio to the form factor $G_{M1}(Q^2)$
\be
\label{eq:fullratio}
R^{\mu}_{\nu}(p', p; \beta, \alpha) = \frac{e \, \varepsilon^{\mu\delta\sigma}{}_{\nu} \, p'_{\delta} \, p_{\sigma} \, G_{M1}(Q^2)}{2\,\sqrt{E_{\pi_{\alpha}}(\p)\,E_{\rho_{\beta}}(\pp) \, \left( (p'_{\nu})^2 - m_{\rho_\beta}^2 g_{\nu\nu} \right)}} \, .
\ee

For our calculation we use SST-propagators evaluated by fixing the current. In particular, we fix the current 3-momentum to $\q = \pm \frac{2\pi}{L} \hat{x} \equiv \pm \vec{\xi}$, the current polarization $\mu=3$ and the vector meson polarization to be $\nu=2$. Using these kinematics, the term $\varepsilon^{\mu\delta\sigma}{}_{\nu} p'_{\delta} p_{\sigma}$ reduces to $(p'_{0} p_{1} - p_{0} p'_{1})$. We choose to refine this further  by taking one of the particles to be at rest. In doing so, we still have the freedom to choose which of the particles is at rest giving rise to the distinct kinematics choices: $\pp = \vec{\xi}$, $\p = 0$ ($\q = + \vec{\xi}$) and $\pp = 0$, $\p = \vec{\xi}$ ($\q = - \vec{\xi}$), each of which gives distinct values of $Q^2$. Applying these kinematics to Eq.~\eqref{eq:fullratio} we arrive at the final expressions used in our evaluation of the $\rho \rightarrow \pi \gamma$ transition form factor
\begin{align*}
	G_{M1}(Q^2) &= \frac{2 \, m_{\rho_\beta}}{e \, |\q|} \sqrt{\frac{E_{\rho_\beta}(\q)}{m_{\pi_\alpha}}} \, R^3_2(\vec{\xi}, 0; \beta, \alpha) \\
	G_{M1}(Q^2) &= \frac{2 \, m_{\rho_\beta}}{e \, |\q|} \sqrt{\frac{E_{\pi_\alpha}(\q)}{m_{\rho_\beta}}} \, R^3_2(0, \vec{\xi}; \beta, \alpha) \, .
\end{align*}

\section{Simulation Details}

For this calculation we use the PACS-CS (2+1)-flavour dynamical-QCD gauge field configurations \cite{Aoki:2008sm} made available through the ILDG \cite{Beckett:2009cb}. These ensembles use a non-perturbatively ${\cal O}(a)$-improved Wilson fermion action and Iwasaki gauge action on a $32^3 \times 64$ lattice, with periodic boundary conditions. The value $\beta=1.9$ provides a lattice spacing of $a = 0.0907$~fm, yielding a physical box length of 2.9~fm. We have access to a total of 5 light quark masses, with the strange quark mass held fixed. The resulting pion masses range from 702~MeV down to 156~MeV. We note that when quoting results away from the physical point we set the scale using the Sommer parameter \cite{Sommer:1993ce} with $r_0 = 0.492$~fm \cite{Aoki:2008sm}. Due to the limited number of configurations at the lighter masses, we make use of multiple quark sources on each configuration. We also evaluate the correlators on the $\lbrace U^* \rbrace$ configurations through the use of the $U^*$-trick outlined in Refs. \cite{Draper:1988xv,Draper:1988bp}.

For the evaluation of the three-point correlators we follow the approach where the SST propagators are evaluated with the current held fixed. For this we use a conserved vector current \cite{Boinepalli:2006xd,Martinelli:1990ny} with polarisation $\mu=3$ and 3-momentum transfer $\q=\pm\frac{2\pi}{L}\hat{x}$. This current is inserted at time $t_c = 21$ relative to the quark source at $t_{src} = 16$. We note that for our choice of ratio we require both $+\q$ and $-\q$, however this is also required for the $U^*$-trick when using SST-propagators. Our error analysis is performed using a second-order jackknife, with the $\chi^2/\mathrm{dof}$ for our fits is obtained through the covariance matrix.

As was done in Ref. \cite{Owen:2015gva}, the anti-quark contribution is evaluated through considerations of charge-conjugation in order to avoid the need to evaluate backwards propagating SST-propagators. In doing so one finds that the quark sectors for the forward and backwards propagating quark fields are of equal magnitude, but differ in sign when exact isospin is present.

For the variational analysis, we use a basis of local meson operators of varying widths \cite{Mahbub:2010rm}. This is achieved by applying different levels of gauge-invariant Gaussian smearing to the quark sources and sinks. By using a variety of widths, the resulting optimised interpolators appear to form nodal structures in the radial wave-function \cite{Roberts:2013oea}. Thus we expect our operators to have greatest overlap with states that contain dominant s-wave components. As these states are strongly peaked at the origin, we would expect these states to have the largest overlap with standard local operators and so the dominant source of excited state contamination for the standard single source approach. The local operators we choose to use are
\begin{align*}
	\chi_{\pi}(x) &= \overline{d}(x) \, \gamma_5 \, u(x) \\
	\chi_{\rho,\,i}(x) &= \overline{d}(x) \, \gamma_i \, u(x) \, ,
\end{align*}
with four different smearing widths, allowing for the construction of a $4 \times 4$ correlation matrix. The correlation-matrix analysis is performed using $t_0 = 17$ with $\delta t = 3$ for the three heavier masses and $\delta t = 2$ for the remaining two lightest masses. 

\section{Results}

\begin{figure}[t]
	\centering
	\includegraphics[width=\columnwidth]{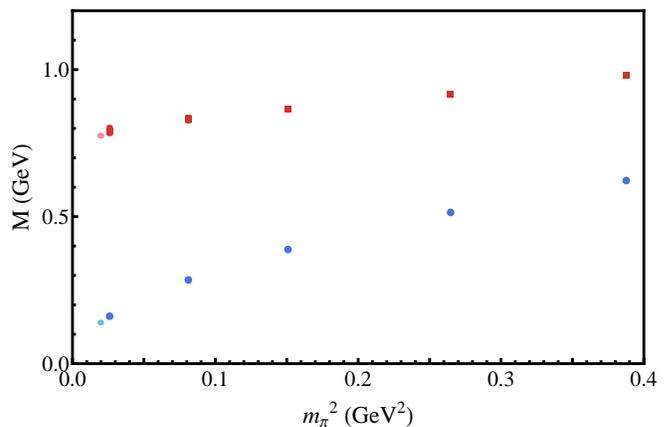}
	\caption{Masses of the $\pi$ (blue) and $\rho$ (red) mesons. The lighter data points at the far left are the corresponding PDG averages \cite{Agashe:2014kda}.}
	\label{fig:spec}
\end{figure}

Before we examine the transition form factor, we examine the isolation of the eigenstates in question.  In Fig.~\ref{fig:spec} we present the masses of the $\pi$ and $\rho$ mesons extracted from our variational analysis.  For the lightest quark mass, which is only slightly heavier than physical, we find that both states agree well with experimental measurements.  Nonetheless, for the rho meson we must take care to ensure that the state we have isolated is in fact the single particle state and not a scattering state.  In Ref. \cite{Mahbub:2013bba} it was found that despite coupling poorly to local operators, scattering states can still populate the projected correlators.  At early times, the weak operator coupling acts to suppress their contribution relative to the dominant single particle state. However if the single particle state is above threshold, at large Euclidean times its contribution is suppressed through the exponential behaviour of the correlator rendering the scattering state the dominant term.  There it was noted that in such a scenario, one could identify a second region of linear behaviour in the logarithm of the projected correlator.  We shall use this signature to identify how far we can sample the correlator and still guarantee single particle dominance.

On these ensembles the $\rho$ meson at rest remains lighter than the nearest $\pi\pi$-scattering state, however this is not the case for non-zero momentum.  In Ref. \cite{Owen:2015gva} we performed a detailed check of our extracted eigenstates to ensure that we had in fact isolated the single particle state, by comparing the extracted energies with the dispersion expectation using the mass obtained with the rho-meson at rest.  This is also compared with the corresponding bare $\pi\pi$-scattering states in order to identify where we should expect to see increased mixing between single and multi-particle states.  In all cases we found that the extracted energies agreed well with the dispersion expectation, particularly for the kinematic arrangement relevant to this work.  Furthermore the decay thresholds were found to lie well away from the extracted energies.

For the particular kinematics and parameters considered in this work, we find that the determinations of $G_{M1}$ with the pion at rest all lie below $Q^2 = 0$, while those with the rho meson at rest lie above.  This gives us values for $G_{M1}$ on either side of $Q^2=0$.  In order to compare with experiment and quark model expectations we require a determination of $G_{M1}(0)$.  To do this we choose to interpolate between our extracted values using a monopole ansatz
\begin{equation}
	\label{eq:VMD}
	G_{M1}(Q^2) = \left( \frac{\Lambda^2}{\Lambda^2 + Q^2} \right) \, G_{M1}(0) \, .
\end{equation}
This choice is motivated by vector meson dominance (VMD) arguments which suggest the form factor should exhibit such a behaviour in the region of low $Q^2$. In these models, the pole mass is identified as the vector meson, i.e. $\Lambda \simeq m_{\rho}$.  In Ref. \cite{Crisafulli:1991pn}, it was found that the VMD hypothesis faired poorly with the data, however we note that this study was conducted with rather large values of $Q^2$ stemming from their small lattice volume.  Contrary to this, the results of Edwards \cite{Edwards:2004xj} which examined the transition over a range of $Q^2$ between 0.02--0.6~GeV${}^2$, display behaviour consistent with this expectation.

\begin{figure}[t]
	\centering
	\subfigure[\ $G^{u}_{M1}$ with the $\pi$ at rest.]{\includegraphics[width=\columnwidth]{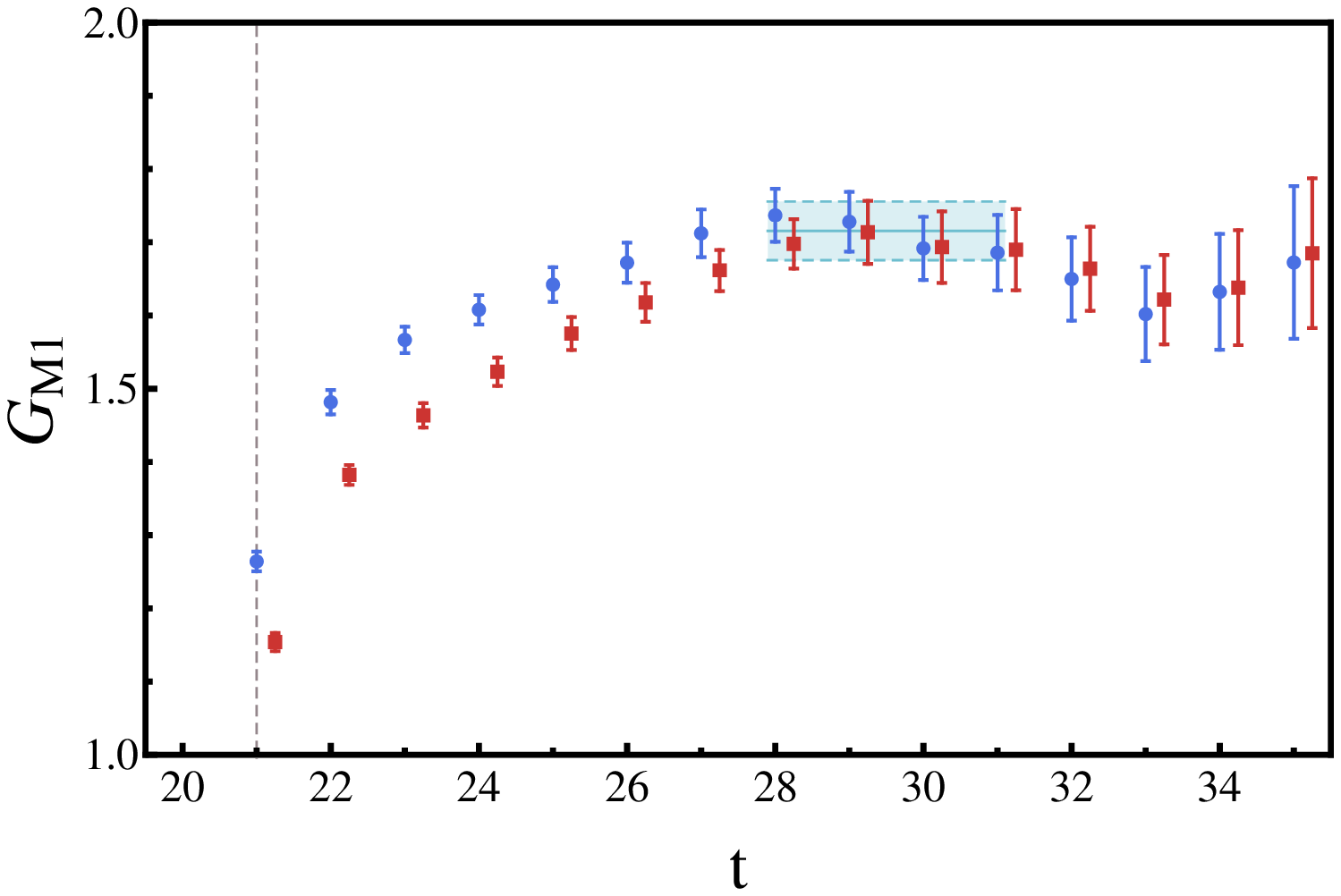}}
	\vspace{3mm}
	\subfigure[\ $G^{u}_{M1}$ with the $\rho$ at rest.]{\includegraphics[width=\columnwidth]{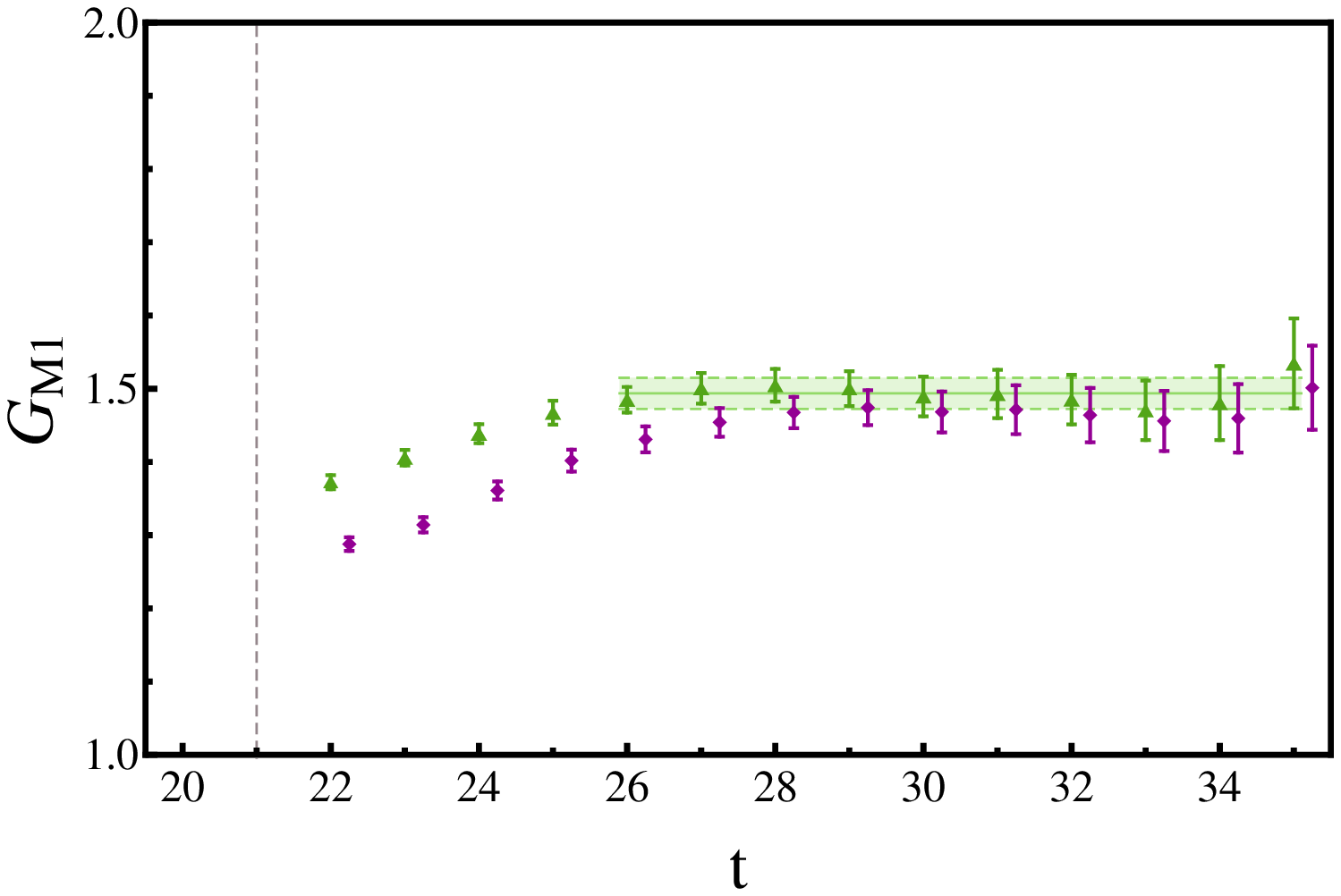}}
	\caption{A comparison of the $u$-quark contribution to the transition form factor, $G_{M1}^{u}$, as a function of Euclidean sink time for a single level of smearing and our variational approach.  The two data sets are offset for clarity.  The upper figure is for the $\pi$ meson at rest with the (blue) circles denoting the results from the variational approach while the (red) squares illustrate traditional results using the standard single-source method with a moderate level of smearing (35 sweeps).  The lower figure is for the $\rho$ meson at rest with the (green) triangles denoting the variational method and the (purple) diamonds denoting the single-source method.  The vertical dashed line indicates the position of the current insertion.  The fitted value from the variational approach has been included (shaded band) to highlight where the single source approach is consistent with our improved method.}
	\label{fig:plat_comp}
\end{figure}

During the extraction of the quark sector contributions to the form factor, we compared the time-series for the ratio using the correlation matrix approach and the standard single smeared source and sink correlator.  For all masses and kinematics, we again find that the correlation matrix method improves the quality of the plateau over modest levels of smearing, however not to the extent observed in previous works \cite{Owen:2012ts,Owen:2015gva}.  In particular, the ratio sampling in the time-like region requires significantly more Euclidean time evolution than the corresponding ratio sampling the space-like region.  We note that in this case the rho meson carries the momentum and the pion is at rest.  Figure~\ref{fig:plat_comp} highlights this comparison for a single quark mass.

\begin{figure}[t]
	\centering
	\includegraphics[width=\columnwidth]{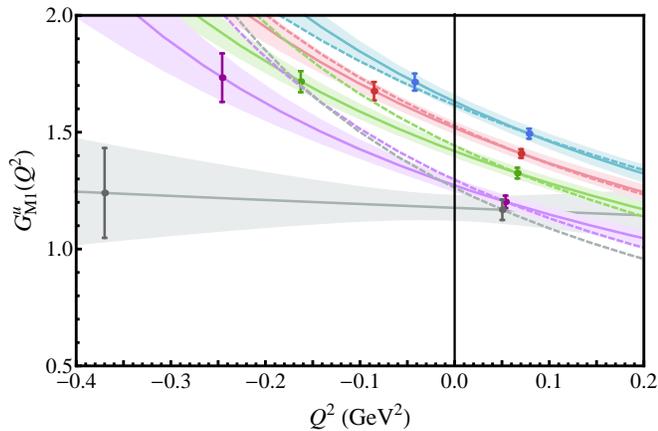}
	\caption{Quark sector results for the transition form factor, $G^{u}_{M1}$. Each curve corresponds to a different value of $m_{\pi}$, with the top curve (blue) corresponding to the heaviest and masses getting lighter as we move down. For each mass we have access to two values of $Q^2$ stemming from the freedom to choose which state is as rest. The solid line and coloured bands are the resulting monopole parametrisation extracted from the two data points, allowing us to access $G_{M1}(0)$. We also include a VMD estimate, dashed line, obtained from using the right positive $Q^2$ data only, with $\rho$-meson mass used as the monopole mass $\Lambda$.}
	\label{fig:GT_w_monopole}
\end{figure}

In Fig.~\ref{fig:GT_w_monopole} we present our results for the extracted values for the $u$-quark sector contribution to the form factor, $G_{M1}^{u}$, as well as the corresponding interpolations used to extract $G^{u}_{M1}(0)$.  Here we choose to label these as the quark sector contributions to the positive-charge eigenstate of the corresponding iso-triplet.  That is, the quark contribution is labelled as the $u$-quark sector while the anti-quark contribution is labelled as the $d$-quark sector. As was mentioned in the previous section, the anti-quark contribution is equal in magnitude with opposite sign, $G_{M1}^{d} = - G_{M1}^{u}$, and so we choose to show the quark contribution only. The quark sector contributions are for quarks of unit-charge. We note that the spread of $Q^2$ sampled is much larger in the time-like region due to the increasing $Q_0$-component stemming from the Goldstone nature of the pion. In the space-like region, we see this effect is suppressed when the momentum is coupled to the pion.  This tight grouping of $Q^2$ values shows a clear decrease in the value of $G_{M1}^{u}$ as the quark masses get lighter.

\begin{figure}[t]
	\centering
	\includegraphics[width=\columnwidth]{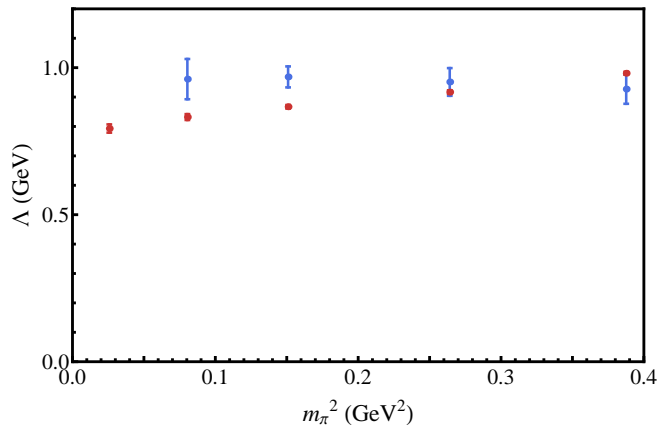}
	\caption{Monopole mass, $\Lambda$, (blue) obtained from applying monopole ansatz to the data. The corresponding $\rho$-meson mass (red) is included to test the VMD hypothesis. The monopole mass for the lightest quark mass is not included due to the large uncertainty in $\Lambda^2$ stemming from the similar values for $G^{u}_{M1}(Q_1^2)$ and $G^{u}_{M1}(Q_2^2)$.}
	\label{fig:mono_mass}
\end{figure}

In the time-like region, the most striking feature is the significantly small value obtained for the lightest mass.  However, examination of the projected two-point function appears consistent with single particle state.  We shall note that the plateau for this particular extraction of $G^{u}_{M1}$ differed in nature to those at heavier masses.  Based upon the $\chi^2_{\mathrm{dof}}$, one is able to fit much earlier, however this is likely the result of a significant increase in the uncertainty of $G^{u}_{M1}$ at early time-slices. Guided by the fit-windows at heavier masses we choose to fit at later times, however it is certainly possible that we simply do not have sufficient statistics and so obtain a value for $G^{u}_{M1}$ that is suppressed.  Another possibility is that we are sampling too far into the time-like region for the VMD hypothesis to hold.  Therefore this result may suggest that this process is suppressed at large time-like momentum transfers.  As a check of the potential impact that this may have on the extracted value of $G^{u}_{M1}(0)$, we compare our results with the VMD estimate obtained using space-like data and the rho meson mass as the monopole mass.  This result is shown as the dashed line in Fig.~\ref{fig:GT_w_monopole}.  In Fig.~\ref{fig:mono_mass} we show the monopole masses extracted from our analysis.  We also include the corresponding rho meson masses and find that they compare reasonably well at large quark masses, but diverge in the light quark regime. Indeed, at the lightest mass $\Lambda \simeq 0$.  This draws the monopole ansatz into question deep in the time-like regime. Thus to connect to experiment we also work with results in the space-like regime ($Q^2 > 0$) and employ VMD.

\begin{table}[t]
	\centering
	\caption{Lattice results for the charge-weighted form factor $G_{M1}$ of $\rho^+ \rightarrow \pi^+ \gamma$ for the two available $Q^2$-values for each pion mass.}
	\begin{ruledtabular}
	\begin{tabular}{cccc}
		$\kappa$ & $m_{\pi}$ (GeV) & $Q^2$ (GeV${}^2$) & $G_{M1}(Q^2)$ \\
		\hline
		13700 & 0.623	& -0.042(5)  & 0.572(12) \\
			    &       &  0.078(3)  & 0.498(7)  \\
		\hline
		13727 & 0.514	& -0.085(5)  & 0.559(13) \\
			    &	      &  0.070(3)  & 0.470(6)  \\
		\hline
    13754 & 0.389	& -0.163(6)  & 0.572(15) \\
	  	    &    		&  0.066(3)  & 0.442(8)  \\
		\hline
		13770 & 0.284	& -0.246(10) & 0.578(35) \\
			    &  	    &  0.054(6)  & 0.401(9)  \\
		\hline
		13781 & 0.161	& -0.370(17) & 0.413(64) \\
			    & 	    &  0.050(9)  & 0.389(15)
	\end{tabular}
	\end{ruledtabular}
\label{tab:GMdata}
\end{table}

\begin{figure}[t]
	\centering
	\includegraphics[width=\columnwidth]{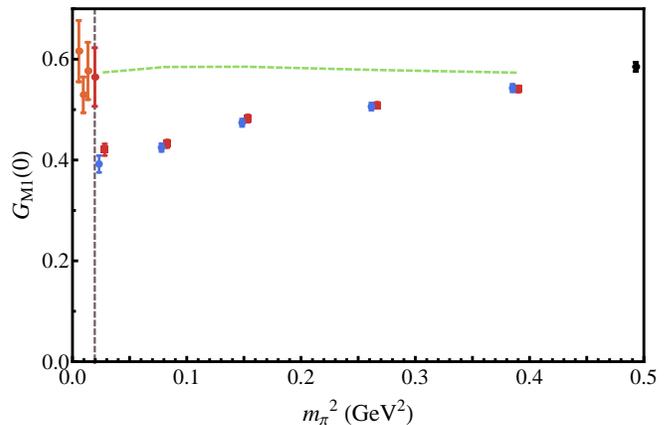}
	\caption{Results for the full hadronic transition moment, $G_{M1}(0)$, extracted using both space and time-like $Q^2$ values (blue circles) and using the space-like values only and invoking the VMD hypothesis (red squares). We include the available experimental extractions (orange) ordered from left to right by year of publication \cite{Jensen:1982nf,Huston:1986wi,Capraro:1987rp}. We also include the PDG average (red) obtained from this data \cite{Agashe:2014kda}. The experimental data are offset for readability, with the PDG average aligned at the physical point indicated by the vertical dashed line. The dashed green line is the non-relativistic quark model expectation of \cite{O'Donnell:1981sj} as discussed herein. We also include the result of Shultz et al. \cite{Shultz:2015pfa} (black) which determines this moment for a single heavy quark mass.}
	\label{fig:GT_hadron}
\end{figure}

We now consider the full hadronic transition form factor.  As was noted earlier, for this transition the quark and anti-quark sector contributions are of equal magnitude and opposite sign and so the charge weighted form factor for $\rho^+ \rightarrow \pi^+ \gamma$ is given by
\begin{align*}
	G_{M1}(Q^2) &= \frac{2}{3} G_{M1}^{u}(Q^2) + \frac{1}{3} (-G_{M1}^{u}(Q^2)) \\ 
	&= \frac{1}{3} G_{M1}^{u}(Q^2) \, .
\end{align*}
Table~\ref{tab:GMdata} summarises our lattice results.  In Fig.~\ref{fig:GT_hadron} we present our results for $G_{M1}(0)$.  We also include the non-relativistic quark model (NRQM) expectation \cite{O'Donnell:1981sj} and the available experimental data.  Due to the different choice of normalisation for this matrix element, we match conventions via the decay width.  The relevant expression for the decay width using our choice of normalisation \cite{Woloshyn:1986pk} is
\be
	\label{eq:decayWidth}
	\Gamma_{\rho\rightarrow\pi\gamma} = \frac{1}{3} \, \alpha \, \frac{|\q|^3}{m_{\rho}^2} \, |G_{M1}(0)|^2 \, ,
\ee
where the photon 3-momentum is evaluated in the rest frame for $\rho$ meson
\[
	|\q| = \frac{m_{\rho}^2 - m_{\pi}^2}{2 m_{\rho}} \, .
\]
Using this expression we are able to evaluate $G_{M1}(0)$ for experimental measurements of the decay width.  We include the PDG average \cite{Agashe:2014kda} as well as the three experimental measurements \cite{Jensen:1982nf,Huston:1986wi,Capraro:1987rp} used in its evaluation.  For the quark model, the choice of normalisation used in Ref. \cite{O'Donnell:1981sj} results in the following expression for the decay width
\[
	\Gamma_{\rho\rightarrow\pi\gamma} = \frac{2}{3} \alpha |\q|^3 \left( \frac{E_{\pi}}{m_{\rho}} \right) \sum_q | \langle \, \rho \, | \frac{\mu_q e_q \sigma_q}{e} | \, \pi \, \rangle |^2 \, .
\]
As discussed in Ref. \cite{O'Donnell:1981sj}, by using SU(6) quark and anti-quark flavour combinations, the sum evaluates to
\[
	\sum | \langle \, \rho \, | \frac{\mu_q e_q \sigma_q}{e} | \, \pi \, \rangle |^2 = \frac{2}{9} \frac{\mu^2_{ud}}{e^2} \, .
\]
Matching with Eq.~\eqref{eq:decayWidth}, this gives rise to the following expression
\[
	G_{M1}(0) = \frac{2}{3} \sqrt{m_{\rho} E_{\pi}} \, \frac{\mu_{ud}}{e} \, .
\]
For the quark moment, $\mu_{ud}$, we use a constituent quark mass that varies linearly in $m_{\pi}^2$
\[
	m_{ud} = a + b \, m_{\pi}^2 \, ,
\]
with $a$ and $b$ fixed such that the constituent quark has a mass of 330~MeV at the physical point and 510~MeV at the SU(3) symmetric point, as determined in Ref. \cite{Agashe:2014kda} using the magnetic moments $\mu_p$, $\mu_n$ and $\mu_{\Lambda}$.
 
Beginning with our results at the heaviest values of $m_\pi$, we find that our lattice data is close to the quark model expectation. Furthermore the observed trend in the data suggests consistency with increasing quark mass where we would expect the quark model expectation to hold. The trend in the data also appears consistent with the determination of Ref. \cite{Shultz:2015pfa}. As we move down to the lighter masses there is a clear downwards trend in the lattice data, as was also observed in our previous quenched study \cite{Owen:2012ej}
\footnote{ We note that in this proceedings, there was an error in converting the experimental decay widths to our definition of the form factor $G_{M1}(0)$. The lattice data however is correct and consistent with results presented herein. Comparison with experiment for \cite{Owen:2012ej} should be done with the value presented here. }.
In contrast to this the quark model result which shows little variation with varying $m_{\pi}$. In the light quark regime, we do not expect the quark model to necessarily hold for this transition. Unlike heavier systems such as bottomium and charmonium decays, the quarks in both the pi and rho meson systems are highly relativistic. Furthermore, the pion itself is a goldstone mode of QCD stemming from the underlying chiral symmetry of the theory. Though we can in principle treat the quarks relativistically \cite{Barik:1992pq,Barik:1994vd}, the inability to properly describe the chiral behaviour of the pion is a fundamental shortfall of all constituent quark models and may certainly lead to deviations in the chiral limit.

Comparison with the experimental determination shows a notable deviation, with the lattice data sitting around 33\% lower than the experimental value.  In Fig.~\ref{fig:GT_hadron} we also include the values obtained using VMD with the space-like data only and find that significant differences persist.  However, we note that our calculation is incomplete.  Unlike the elastic form factors for which disconnected contributions are necessarily zero \cite{Dudek:2006ej,Draper:1988bp}, such contributions are present for meson transitions that involve a change in $G$-parity \cite{Shultz:2015pfa}. This is related to the invariance of the QCD action under charge-conjugation.  For disconnected contributions involving the electromagnetic current, under charge-conjugation the ``$C$-even" two-point function and the ``$C$-odd" disconnected loop give rise to a relative sign between the $\lbrace U \rbrace$ and $\lbrace U^* \rbrace$ configurations and so cancel exactly \cite{Draper:1988bp} in the ensemble average.  For transitions involving a change in $G$-parity, the two-point function is now ``$C$-odd" and so combining with the disconnected loop gives rise to a common sign between the $\lbrace U \rbrace$ and $\lbrace U^* \rbrace$ configurations resulting in a non-zero quantity \cite{Dudek:2006ej,Shultz:2015pfa}.  Furthermore, if one neglects disconnected $s$-quark contributions, the charge weight factors between the connected contributions and the disconnected contributions are equal
\begin{align*}
	G_{M1} &= q_u \, G_{M1}^{\textrm{con.}} + q_{\bar{d}} \, \left( - G_{M1}^{\textrm{con.}} \right) + q_u \, G_{M1}^{\textrm{dis.}} + q_d \, G_{M1}^{ \textrm{dis.}} \\
	&= \left( \frac{2}{3} - \frac{1}{3} \right) \, G_{M1}^{\textrm{con.}} + \left( \frac{2}{3} - \frac{1}{3} \right) \, G_{M1}^{\textrm{dis.}} \\
	&= \frac{1}{3} \, G_{M1}^{\textrm{con.}} + \frac{1}{3} \, G_{M1}^{\textrm{dis.}} \, .
\end{align*}
Thus the discrepancy between our results and the experimental value suggest that disconnected contributions are likely to play an important role in fully describing this transition. One would also expect such contributions to become increasing important with decreasing quark mass. This expectation complements the observation that our results are consistent with quark model expectations at heavier masses, however deviate as we move to light-quark regime.

Another important aspect that warrants further consideration is the fact that the rho meson is ultimately a resonant state.  The nature of resonant states on the lattice is significantly different to that of continuum due to the lack of a continuous distribution of momentum modes and so to properly make connection with the continuum expectation, one must suitably evolve the lattice determinations to the infinite volume to properly account for the differences in the underlying multi-particle interactions.  Understanding exactly how to do this in general is an area of current interest to the community and only very recently has a framework been presented to handle $1 \rightarrow 2$ body processes required to properly describe this transition \cite{Bernard:2012bi,Agadjanov:2014kha,Briceno:2014uqa,Briceno:2014uqa}.  In fully addressing this aspect, it is important to also consider the role of $\pi\pi$ scattering states in the $\rho$-meson correlation matrix to ensure a complete isolation of QCD eigenstates on the finite volume.

\section{Concluding Remarks}

In this work we present the first light quark examination of the radiative decay of the rho meson.  The calculated transition moment, $G_{M1}(0)$ was found to be consistent with quark model expectations at heavy masses.  However we have discovered an important quark mass dependence.  Our results in the light quark regime sit low in comparison with experimental determinations, suggesting important disconnected contributions to this process.

These results warrant a more comprehensive investigation of this process.  Any future work should aim to focus on the inclusion of disconnected contributions, multi-particle contributions and finite volume effects so as to allow for the proper evolution of the lattice determination to the infinite-volume.  Understanding the role these systematics play in this calculation may provide important insights into  topical transition amplitudes such as $N^* \rightarrow N \gamma^*$ and $\Delta^{(*)} \rightarrow N \gamma^*$ transitions, and those central to the search for exotic hadron states \cite{Dudek:2012vr}.

\section*{Acknowledgements}

\sloppy We thank PACS-CS Collaboration for making their $2+1$ flavor configurations available and acknowledge the important ongoing support of the ILDG. This research was undertaken with the assistance of resources at the NCI National Facility in Canberra, Australia, and the iVEC facilities at the Pawsey Supercomputing Centre, Murdoch University (iVEC@Murdoch) and the University of Western Australia (iVEC@UWA). These resources were provided through the National Computational Merit Allocation Scheme, supported by the Australian Government. We also acknowledge eResearch SA for their support of our supercomputers. This research is supported by the Australian Research Council through Grant No. DP120104627 (D.B.L.).





\bibliography{mesFF_trans}


\end{document}